\documentclass[dvips,12pt]{article}
\usepackage{hyperref}
\usepackage{amsfonts}
\usepackage{amsmath, amssymb}
\usepackage{graphicx}
\usepackage{psfrag}
\usepackage{youngtab}
\Yboxdim{5pt}

\newcommand{\ra}{\rightarrow}
\newcommand{\la}{\leftarrow}

\newcommand{\be}{\begin{equation}}
\newcommand{\ee}{\end{equation}}
\newcommand{\ba}{\begin{eqnarray}}
\newcommand{\ea}{\end{eqnarray}}
\newcommand{\bi}{\begin{itemize}}
\newcommand{\ei}{\end{itemize}}

\newcommand{\Tr}{{\rm Tr}}

\newcommand{\Z}{{\bf Z}}
\newcommand{\R}{{\rm R}}

\newcommand{\p}{\partial}

\newcommand{\Ncal}{{\mathcal N}}

\newcommand{\Kahler}{K\"{a}hler }

\newcommand{\nn}{\nonumber}
\newcommand{\mo}{{-1}} 

\newcommand{\f}{\frac}
\newcommand{\half}{\frac{1}{2}}
\newcommand{\oo}{\frac{1}}

\newcommand{\Path}{{\rm P}}
\def\Dslash{\,\,{\raise.15ex\hbox{/}\mkern-12mu D}}
\def\Dbarslash{\,\,{\raise.15ex\hbox{/}\mkern-12mu {\bar D}}}
\def\delslash{\,\,{\raise.15ex\hbox{/}\mkern-9mu \partial}}
\def\delbarslash{\,\,{\raise.15ex\hbox{/}\mkern-9mu {\bar\partial}}}
\def\pslash{\,\,{\raise.15ex\hbox{/}\mkern-9mu p}}
\def\calDslash{\,\,{\raise.15ex\hbox{/}\mkern-12mu {\cal D}}}

\renewcommand{\bar}{\overline}
\renewcommand{\tilde}{\widetilde}
\renewcommand{\hat}{\widehat}

\begin{document}
\baselineskip=15.5pt
\renewcommand{\theequation}{\arabic{section}.\arabic{equation}}
\pagestyle{plain}
\setcounter{page}{1}
\bibliographystyle{utcaps}
\begin{titlepage}

\rightline{\small{\tt NSF-KITP-07-109}}
\begin{center}

\vskip 3 cm

{\Large {\bf  BIons in topological string theory}}

\vskip 3cm
Takuya Okuda

\vskip 1cm

Kavli Institute for Theoretical Physics

University of California,
Santa Barbara

CA 93106, USA

\vskip 3cm

{\bf Abstract}

\end{center}

When many fundamental strings are stacked together, they puff up into D-branes.
BIons and giant gravitons are the examples of such D-brane configurations
that arise from coincident strings.
We propose and demonstrate analogous transitions in topological string theory.
Such transitions can also be understood in terms of
the Fourier transform of  D-brane amplitudes.


\end{titlepage}

\newpage

\section{Introduction and summary}

A fascinating aspect of string theory is duality,
the equivalence of seemingly different descriptions of a physical system.
In some classes of duality, physical objects placed in a region of spacetime
have a description
in terms of different kind of objects.
An important example is geometric transition:
when many D-branes are placed on top of each other,
the system is better described by a new geometry with no D-branes.
Study of geometric transitions has led to many important insights and results,
including AdS/CFT correspondence \cite{Maldacena:1997re}, microscopic explanation of black hole entropy \cite{Strominger:1996sh},
and the relation between gauge theories and matrix models \cite{Dijkgraaf:2002fc,Dijkgraaf:2002dh}.

Another class of such ``local'' duality is the transition of fundamental
strings to D-branes (we will call it the {\it string/brane transition}),
in which a  system of many coincident strings 
can be described by D-branes that replace the strings.
As an example, let us consider the BIon solution 
\cite{Callan:1997kz,Gibbons:1997xz}.
When many fundamental strings end on a D-brane,
the D-brane world-volume backreacts
and develops a spike, which sticks out in the direction of the strings.
Such embedding of the world-volume to spacetime
is a solution to the equations of motion
of the Born-Infeld action, and it is a dual description of the strings.
Another early example appeared in  AdS/CFT correspondence:
the Kaluza-Klein graviton modes (fundamental strings that 
rotate in the 5-sphere) puff up into
giant gravitons (D-branes that wrap a 2-sphere and rotate in the 5-sphere)
 \cite{McGreevy:2000cw,Grisaru:2000zn,Hashimoto:2000zp}.
A newer example of string/brane transition was found in the study of
Wilson loops in AdS/CFT.
The Wilson loop in the fundamental representation
is described by 
a string world-sheet that
extends in the bulk of $AdS_5$ and ends along the loop on the boundary \cite{Maldacena:1998im}.
For higher representations, 
the Wilson loops 
have bulk realizations
in terms of D3- or alternatively D5-branes
\cite{Drukker:2005kx, Yamaguchi:2006tq, Gomis:2006sb, Gomis:2006im}.
Such D-branes are the dual descriptions of many fundamental strings.

\begin{figure}[htbp]
\centering
\begin{tabular}{cccccc}
\psfrag{S3}{$S^3$}
\includegraphics[scale=.45]{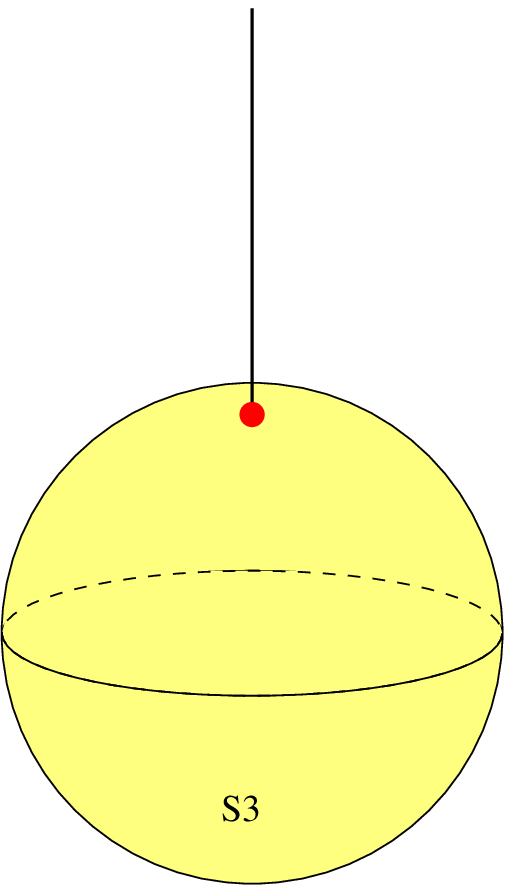}

\hspace{10mm}
&
\psfrag{S3}{$\R^2\times S^1$}
\includegraphics[scale=.45]{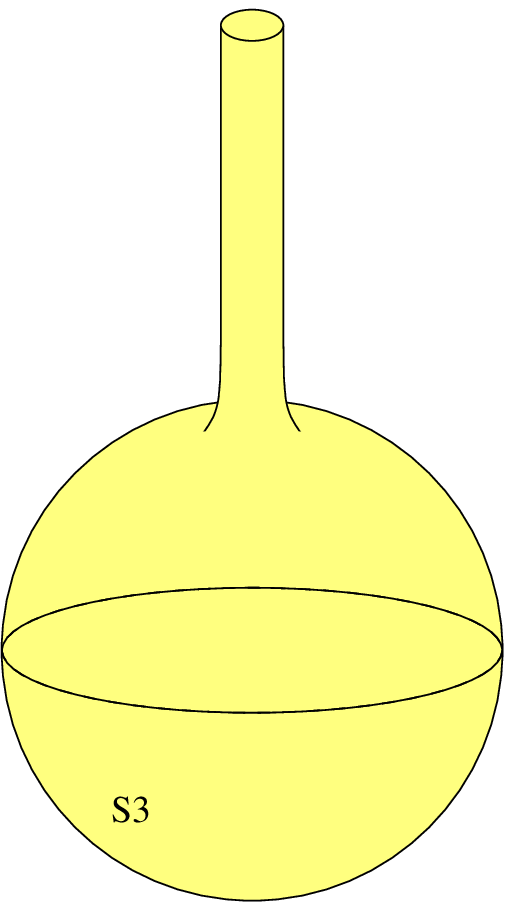}
\\
(a)&&(b)
\end{tabular}
\caption{
A BIon in the deformed conifold (an artist's impression).
(a) Fundamental strings end on D-branes wrapping the
$S^3$.  The end points are along a knot suppressed in the figure.
(b) D-branes develop a ``spike'' and become non-compact with topology $\R^2\times S^1$.
}
\label{BIon1}
\end{figure}

In this paper, we propose similar transitions in topological string theory.
An illuminating example is the analog of a BIon.
Consider the deformed conifold (the total space of $T^* S^3$)
and wrap $P$ D-branes around the $S^3$.
Let many  non-compact string world-sheets 
end on the branes along a knot,
and suppose they 
extend in the fiber direction as shown in figure \ref{BIon1}a.
This system is equivalently described by 
$P$ non-compact D-branes of topology ${\rm R}^2\times S^1$ without strings\footnote{
The equivalent transition in the resolved conifold was discussed in \cite{Gomis:2006mv},
and is revisited in Appendix \ref{st-br-identity-section},
where a more elementary calculation is presented.
}, as in  figure \ref{BIon1}b.
We  motivate the string/brane transitions in two ways.

\begin{figure}[ht]
\centering
\psfrag{a}{Fundamental}
\psfrag{a2}{string}
\psfrag{b}{D-brane}
\psfrag{c}{Anti-brane}
\psfrag{d}{Bubbling Calabi-Yau}
\includegraphics[scale=.45]{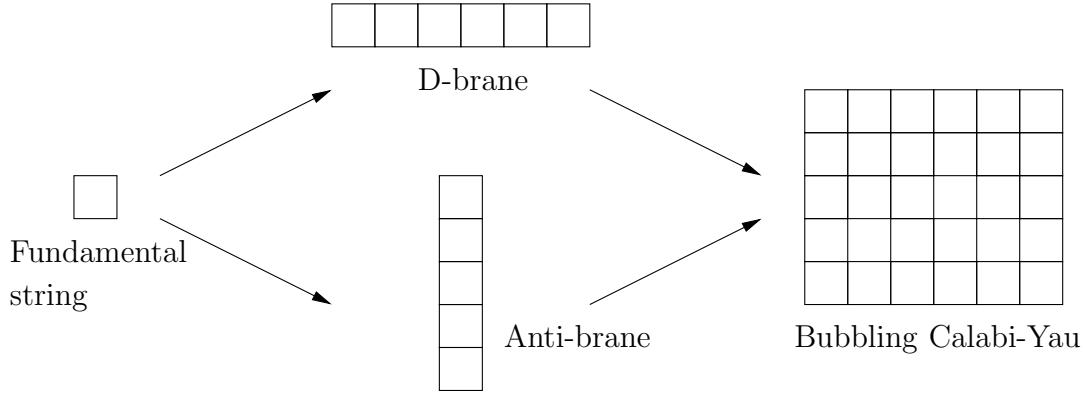}
\caption{Gravity duals of a Wilson loop in Chern-Simons theory.
A string corresponds to the fundamental representation,
a D-brane to a symmetric representation,
an anti-brane to an anti-symmetric representation,
and a bubbling Calabi-Yau to the representation specified 
by a rectangular Young tableau \cite{Gomis:2006mv,Gomis:2007kz}.
For a local operator in $\Ncal=4$ Yang-Mills,
replace (D-brane, Anti-brane) in the figure by (D3-brane wrapping $S^3\subset AdS_5$, D3-brane 
wrapping $S^3\subset S^5$)
 \cite{McGreevy:2000cw,Grisaru:2000zn,Hashimoto:2000zp,
Balasubramanian:2001nh, Corley:2001zk}.
For a Wilson loop in the same theory, 
replace it by (D3-brane wrapping $S^2\subset AdS_5$, 
D5-brane wrapping $S^4\subset S^5$)
 \cite{Drukker:2005kx, Yamaguchi:2006tq, Gomis:2006sb, Gomis:2006im}.
For the both operators in Yang-Mills,
replace (bubbling Calabi-Yau)
by (bubbling supergravity solution)
\cite{Lin:2004nb,Yamaguchi:2006te,Lunin:2006xr}.
}
\label{young-trans}
\end{figure}

First, these  transitions play a role in
gauge/gravity correspondence
 \cite{Maldacena:1997re,Gopakumar:1998ki}.
There is a universal pattern in the correspondence 
between 
operators of the form $\Tr_R(\ldots)$ in gauge theory,
and their dual objects in gravity.
These operators are labeled by a Young tableau $R$,
which specifies a representation of $U(N)$.
Each box corresponds to a fundamental string,
a single row  to a  D-brane,
a single column to another type of D-brane,
and a large rectangle to a new cycle in geometry. 
With
Wilson loops in Chern-Simons theory taken as an example,
the transition of strings to branes 
and the further transition of branes to geometry
are summarized in figure \ref{young-trans}.\footnote{
\label{convention}
We choose the convention so that a anti-brane 
here is a D-brane (rather than an anti-brane) in \cite{Gomis:2006mv} and 
vice versa,
for reasons explained in footnote \ref{convention-reason}.}
This paper studies the transition of strings to branes
 in more general settings.
The related reference \cite{Gomis:2007kz} focused on the general transition
of branes to geometry.

Second, another line of development
found
that D-brane amplitudes in topological string theory are wave functions
of Chern-Simons theory \cite{Aganagic:2003qj}.
The relevant D-branes are non-compact, and
 a state in the Hilbert space specifies 
the boundary condition at infinity.
As in ordinary quantum mechanics,
one performs the Fourier transform
of wave functions when we change the basis.
In an appropriate basis,
the boundary condition is consistent with
two descriptions that are seemingly different.
In the first description, the boundary condition
picks out a specific configuration
of  strings ending on the branes.
We can regard the branes as an effective cut-off to the
infinite world-sheets, and we interpret the system as
the configuration extended strings.
In the second description, the same boundary condition 
specifies a particular value of the holonomy carried by the branes.
Consistency requires the equivalence of the two descriptions.


The rest of the paper assumes the
knowledge of topological strings
in toric Calabi-Yau manifolds
\cite{Aganagic:2003db}.
In section \ref{st-br-toric-section},  we propose the transitions
of strings to D-branes
in such  geometries.
We prove
the equality
of partition functions in the two descriptions
as evidence for the proposal.
Next, section \ref{BIon-section} discusses the analog of BIons.
This is the large $N$ dual reinterpretation
of the string/brane transitions in section \ref{st-br-toric-section}.
We propose that
fundamental strings ending on compact D-branes  are
dual to non-compact D-branes, 
and verify the proposal by
matching their partition functions.
Finally in section \ref{alternative-section},
we explain that 
a certain boundary condition
admits two characterizations, one in terms of a string configuration
and the other in terms of holonomy of the branes.
Their compatibility is a physical explanation of the string/brane transition.
In Appendix \ref{rederivation} we present an elementary derivation
of the equality between
 the unknot Wilson loop vev and a D-brane amplitude.
We also formulate a conjecture that relates
knot polynomials and D-brane amplitudes for general knots.
Other appendices explain the formalisms and calculations that are used in the main text.

\section{String/brane transitions in toric Calabi-Yau manifolds}
\label{st-br-toric-section}

As a basic example of the transition of strings to D-branes,
we consider gravity duals of Wilson loops in Chern-Simons theory.
If the Young tableau $R$ has $P$ rows,
the Wilson loop $\Tr_R \Path e^{-\oint A}$ is
dual to 
$P$ D-branes in the resolved conifold
 with holonomy determined by $R$ \cite{Gomis:2006mv}.
The Wilson loop is a linear combination of multi-trace operators
and each single trace is dual to a fundamental string
wrapping a non-compact holomorphic surface.
Therefore {\it the D-brane configuration is a superposition 
of multi-string states}, just as argued for giant gravitons 
in \cite{Balasubramanian:2001nh}.
The Frobenius relation
\ba
\Tr_R U=\sum_{\vec k} \oo{z_{\vec k}} \chi_R(C(\vec k))\Tr_{\vec k} U
\ea
tells us how to superpose the  multi-string states.
Here $C(\vec k)$ is the conjugacy class of 
the symmetric group $S_k$ specified
by the partition $\vec k=(k_1,k_2,\ldots)$ of $k=\sum_j j k_j$,
and the symbol $\chi_R(C(\vec k))$ denotes the character.
We have also defined $z_{\vec k}\equiv \prod_j k_j! j^{k_j}$ and
$\Tr_{\vec k} U\equiv \prod_j (\Tr U^j)^{k_j}$.
Let ${\rm F1}_{\vec k}$ be the state that has
$k_j$ strings wrapping the holomorphic surface $j$ times for all 
positive integers $j$.
Then the superposition
\ba
{\rm F1}_R\equiv \sum_{\vec k} \oo{z_{\vec k}} \chi_R(C(\vec k))
{\rm F1}_{\vec k}
\ea
is dual to the D-branes.

\begin{figure}[ht]
\centering
\begin{tabular}{cccccc}
\psfrag{R}{${\rm F1}_R$}
\psfrag{R2}{$R_2$}
\psfrag{R3}{$R_3$}
\psfrag{t2}{$t_2$}
\psfrag{t3}{$t_3$}
\includegraphics[scale=.45]{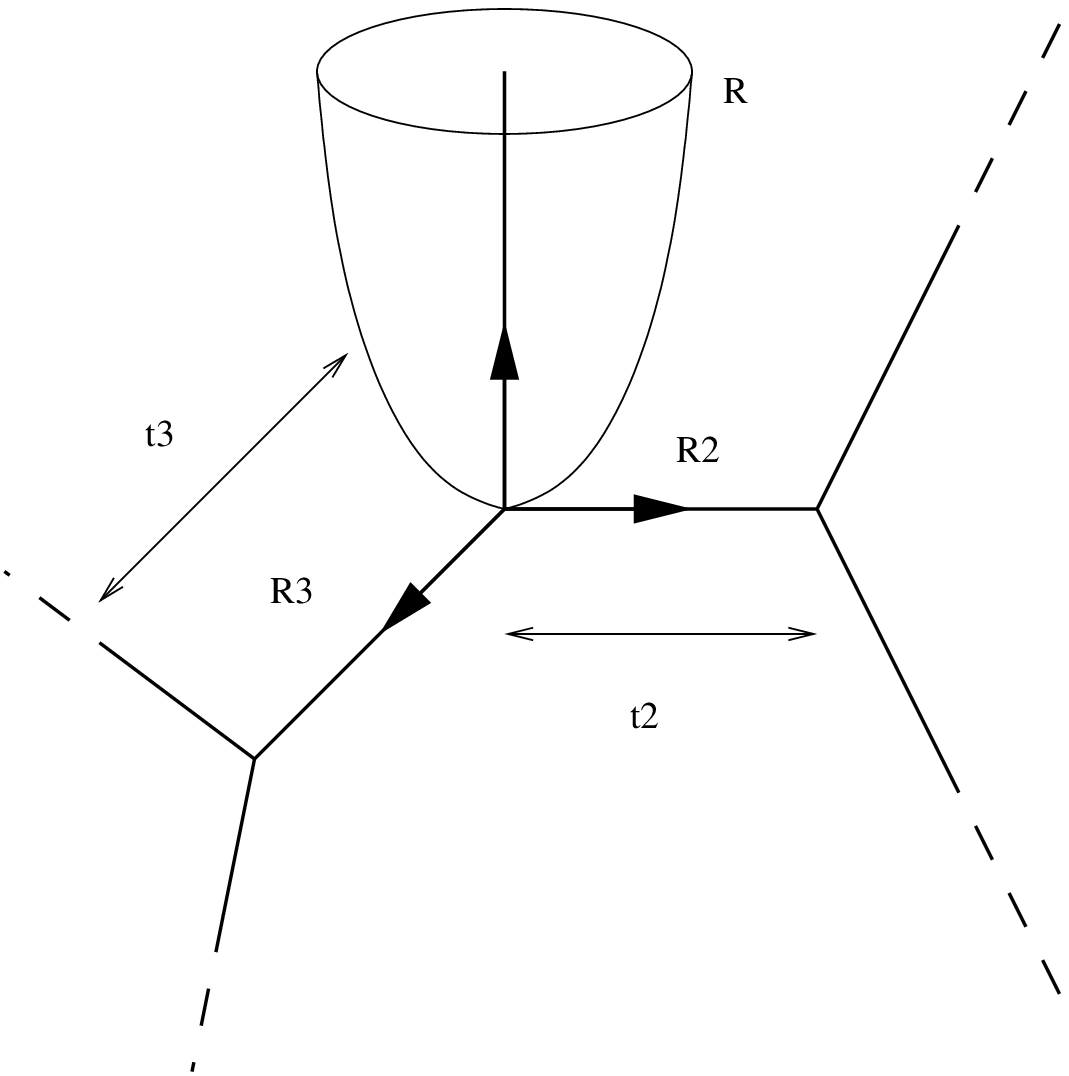}
&
\psfrag{f}{}
\psfrag{R3}{$R_3$}
\psfrag{R2}{$R_2$}
\psfrag{a1}{$a_1$}
\psfrag{aM}{$a_P$}
\psfrag{Q2}{$Q_2$}
\psfrag{Q2prime}{$Q$}
\psfrag{t2}{$t_2+g_sP$}
\psfrag{t3}{$t_3-g_sP$}
\includegraphics[scale=.45]{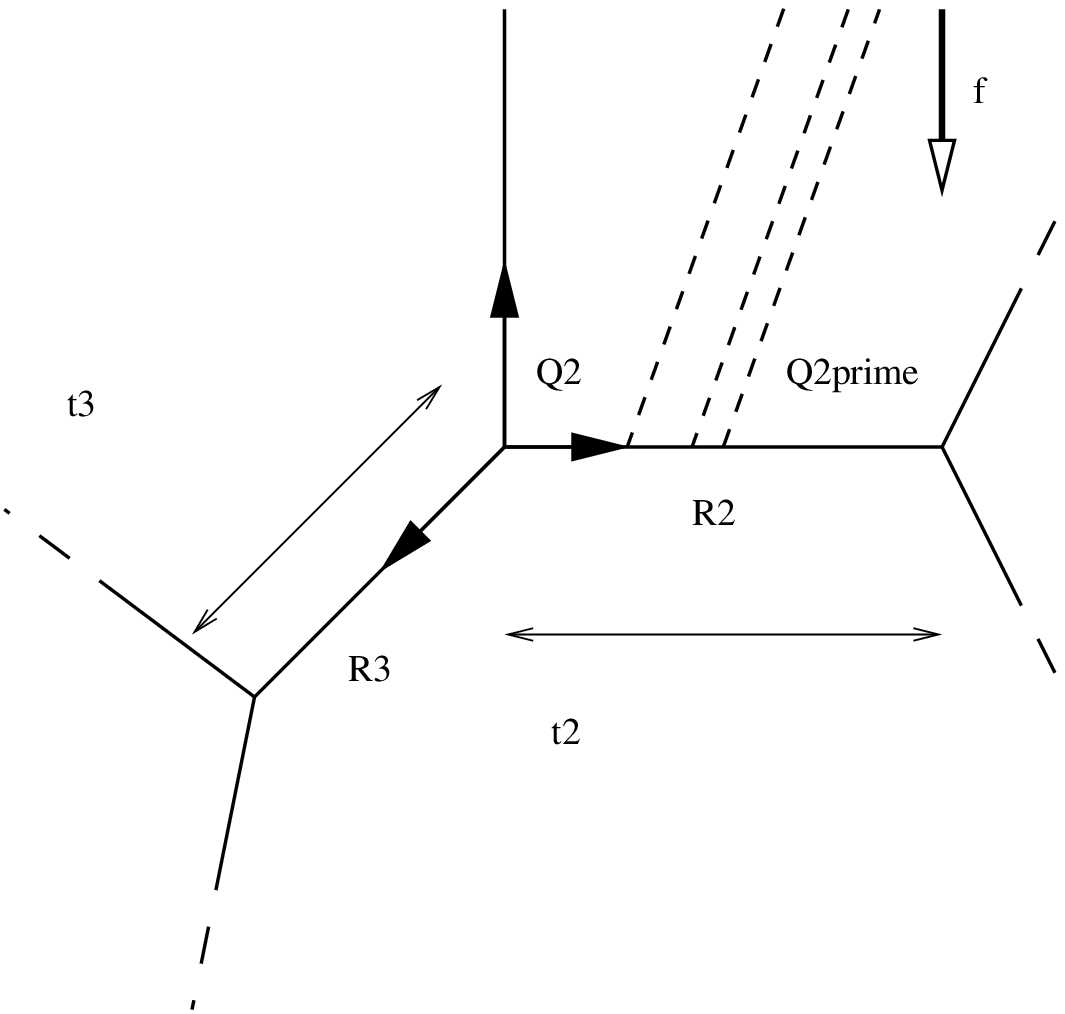}
&
\psfrag{f}{}
\psfrag{R3}{$R_3$}
\psfrag{R2}{$R_2$}
\psfrag{a1}{$a_1$}
\psfrag{aM}{$a_P$}
\psfrag{Q3}{$Q_3$}
\psfrag{Q3prime}{$Q$}
\psfrag{t2}{$t_2-g_sM$}
\psfrag{t3}{$t_3+g_sM$}
\includegraphics[scale=.45]{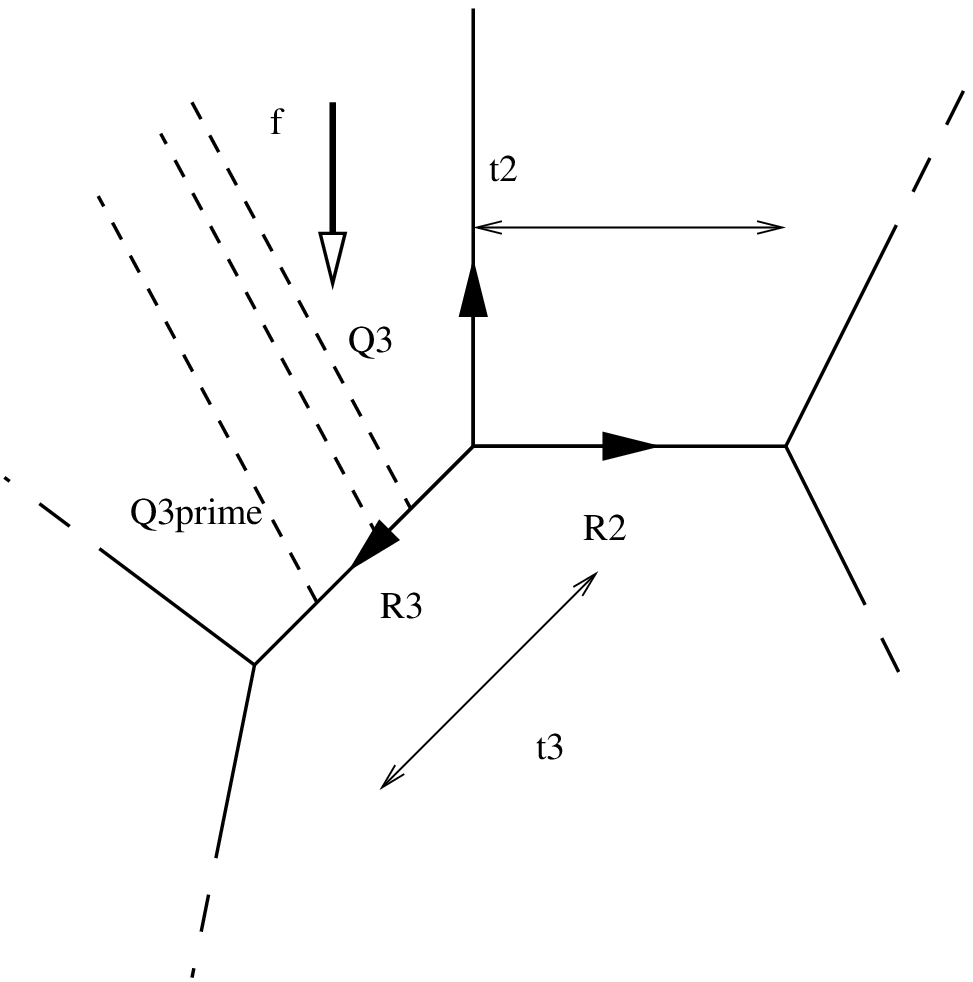}
\\
(a)&(b)&(c)
\end{tabular}
\caption{
(a) Non-compact string world-sheets in a toric Calabi-Yau manifold.
(b) The strings puff up into the D-branes
represented by the dashed lines.
The holonomy is gauge equivalent to the positions of branes.
(c) The strings are replaced by anti anti-branes.
}
\label{string-brane}
\end{figure}

For generalization, let us now consider an arbitrary toric Calabi-Yau manifold
specified by a  web diagram.
A semi-infinite edge represents a non-compact cycle of topology ${\rm R}^2$,
and shares a vertex with two other edges
as shown in figure \ref{string-brane}a.
Let us consider the configuration ${\rm F1}_R$ of
strings wrapping the cycle as defined above.
We propose that these strings are dual to  
$P$ D-branes inserted at a neighboring
edge (figure \ref{string-brane}b).
As in \cite{Gomis:2006mv},
the distance of the $i$-th brane from the vertex is 
\ba
y_i\equiv g_s(R_i-i+P+1/2),
\ea
where $R_i$ is the number of boxes in the $i$-th row.
Generalizing the anti-brane realization of Wilson loops
\cite{Gomis:2006mv},
we also propose that the  strings are dual to 
$M$ anti-branes at another neighboring edge
(figure \ref{string-brane}c).
$M$ is the number of columns in $R$, and the $i$-th anti-brane
is distance $x_i\equiv g_s(R^T_i-i+1/2+M)$ away from the  vertex.


We now provide quantitative evidence for our proposals,
by showing the
equality of the partition functions in the three descriptions.
How do we define the partition function
for the fundamental strings?
We define the function $Z_{\vec k}$
as the sum of all world-sheet diagrams
that share the asymptotics ${\rm F1}_{\vec k}$.
The asymptotics requires us to
include $k_j$ world-sheets that are wrapped $j$ times for 
all $j$.
Since  the world-sheets are non-compact, we regularize
the values of the diagrams by subtracting the infinite area.
We then define the partition function for the configuration ${\rm F1}_R$
as
\ba
Z_R\equiv \sum_{\vec k} \oo{z_{\vec k}} \chi_R(C(\vec k))
Z_{\vec k}.
\ea

The partition functions $Z_R$  and $Z_{\vec k}$ for strings
are related to the partition function $Z(V)$ 
in the presence of D-branes with holonomy $V$ as
\ba
Z(V)=\sum_R Z_R\Tr_R V=\sum_{\vec k}\oo{z_{\vec k}} Z_{\vec k}\Tr_{\vec k}V,
\ea
because $Z(V)$ is defined as a sum over all string configurations.\footnote{
Section \ref{alternative-section} studies
the  relation between $Z_R$ and
$Z(V)$ in more detail.}
Because D-brane amplitudes can be computed using the topological vertex  
$C_{R_1 R_2 R_3}(q)$ \cite{Aganagic:2003db},
so can the partition functions for the strings.
We find that\footnote{
Our convention is such that $q\ra q^\mo$ relative to \cite{Aganagic:2003db}.
Appendix  \ref{top-vert-app} summarizes useful formulas.
}
\ba
Z_R=\sum_{R_2,R_3}C_{R R_2 R_3}(q) e^{-|R_2|t_2}e^{-|R_3|t_3} \times \ldots, 
\label{st-eq1}
\ea
where we have defined $q\equiv e^{-g_s}$ and $g_s$ is the 
string coupling constant.
$|R_a|$ denotes the number of boxes in the Young tableau $R_a$.
The \Kahler moduli $t_2$ and $t_3$ are 
defined in figure \ref{string-brane}a.
The topological vertex amplitude
 $C_{R R_2 R_3}$ represents 
the contribution from the vertex at the center of figure \ref{string-brane}a.
We focus on it because all the rest in $Z_R$ is not affected
in the transitions.

For the test of string/D-brane transition,
the key identity is\footnote{
For the reason explained in section \ref{alternative-section},
it is slightly more natural to consider $q^{\half \kappa_R} C_{R R_2 R_3}$.
When we apply (\ref{st-br-identity}) to
$q^{\half \kappa_R} C_{R R_2 R_3}$,
$R^T$ does not appear.
}
\ba
C_{R R_2 R_3}&=&
\xi(q)^{P}
\prod_{1\leq i<j\leq P} (1-e^{-(y_i-y_j)})
e^{ |R_3|g_s P-|R_2|g_sP}
 q^{-\half \kappa_{R_2}+\half ||R^T||^2 }
\nn\\
&&\times\sum_{Q_2,Q} C_{\cdot Q_2 R_3}(-1)^{|Q_2|} q^{\half \kappa_{Q_2}}
\Tr_{Q_2/Q} U_R (-1)^{|Q|} \Tr_{R_2{}^T/Q{}^T} U_R{}^\mo. \label{st-br-identity}
\ea
Let us explain the notation.
We have defined $\xi(q)\equiv 1/\prod_{j=1}^\infty (1-q^j)$, 
$\kappa_R\equiv \sum_i (R_i-2i+1)R_i$, 
$||R||^2\equiv \sum_i R_i^2$, and
$U_R\equiv {\rm diag}(e^{-y_i})_{i=1}^P$.
$R^T$ is the transposed diagram.
The symbol $\Tr_{R/Q}(X)$ denotes the skew Schur polynomial $s_{R/Q}$
(\ref{skew})
whose arguments are the eigenvalues of the matrix $X$.
We prove the identity (\ref{st-br-identity}) 
in Appendix \ref{st-br-identity-section}.

After we apply the identity (\ref{st-br-identity}) to 
(\ref{st-eq1}), $Z_R$ 
 becomes the partition function for
$P$ D-branes at distances $y_i$ from the vertex
(figure \ref{string-brane}b).
We can see this as follows.
According to the gluing rules of \cite{Aganagic:2003db},
the second line  of (\ref{st-br-identity}) 
indicates that $P$ D-branes are inserted  at distances $y_i$.
The holomorphic annuli 
between the $i$-th and $j$-th D-branes
contribute the factor $(1-e^{-(y_i-y_j)})$.
The extra exponential
$e^{ |R_3|g_s P-|R_2|g_sP}$ tells us that the edge associated with $R_2$ grows
in size by $g_sP$
while the one associated with $R_3$ shrinks by the same amount.
An arrow in the figure indicates the framing
that we read off from the factor $q^{-\half \kappa_{R_2}}$.\footnote{
The function $\xi(q)^P$ is not important in perturbation theory,
and no clear interpretation is known.
The factor $q^{\half ||R^T||^2 }$  combined with $\prod(1-e^{-(y_i-y_j)})$
makes the whole expression anti-symmetric in $y_i$, exhibiting the
fermionic nature of the non-compact D-branes.}

We can also use the following identity
to verify the string/anti-brane transition:\footnote{
The proof is completely parallel to the proof of (\ref{st-br-identity}) and is omitted.
}
\ba
 C_{RR_2 R_3}&=&
\xi(q)^{M} \prod_{1\leq i<j\leq M} (1-e^{-(x_i-x_j)})  q^{-M|R_2|+M|R_3|}
 q^{\half ||R^T||^2}
\sum_{Q_3,Q}
C_{\cdot R_2Q_3}(-1)^{|Q_3|} \nn\\
&&\times
\Tr_{{Q_3}^T/Q} U_{R^T}(-1)^{|Q|} \Tr_{R_3/Q^T} U_{R^T}^\mo.
\ea
Here
$U_{R^T}\equiv{\rm diag}(e^{-x_i})$.
The rules for inserting anti-branes are summarized in section 4 
of \cite{Gomis:2007kz}.
After we apply the identity to (\ref{st-eq1}), $Z_R$
becomes the partition function for 
the anti-branes in figure \ref{string-brane}c.
The size of the edge  with anti-branes 
increases by $g_sM$, and
the adjacent edge shrinks as much.
Figure \ref{string-brane}c also shows the framing of the anti-branes.

To summarize, we have shown the equality of 
the partition functions in three descriptions, namely those in terms
of strings, D-branes, and anti-branes.
This section has mostly focused on the quantitative evidence,
and section \ref{alternative-section} will give a physical explanation
of the transitions.
Before doing that, in the next section we will
deal with a large $N$ dual reinterpretation
of the string/brane transitions.

\section{BIons}\label{BIon-section}

Let us consider a D-brane in physical string theory with
many transverse fundamental strings 
ending on it.
This system has a dual description as a solution to the
equations of motion of the Born-Infeld  action
\cite{Callan:1997kz,Gibbons:1997xz}.
The brane world-volume has a spike poking out
in the transverse direction, and the electric flux supports the
non-trivial profile.
The spike has replaced the coincident strings, and
there non-zero gauge flux 
as the remnant of  string charge.
Such a brane solution is known as a BIon \cite{Gibbons:1997xz},
and we propose
its analog in topological string theory.
We will begin with a basic example and generalize it in two steps.

\begin{figure}[ht]
\centering
\begin{tabular}{cccc}
\psfrag{Compact}{$P$}
\psfrag{R}{${\rm F1}_R$}
\includegraphics[scale=.45]{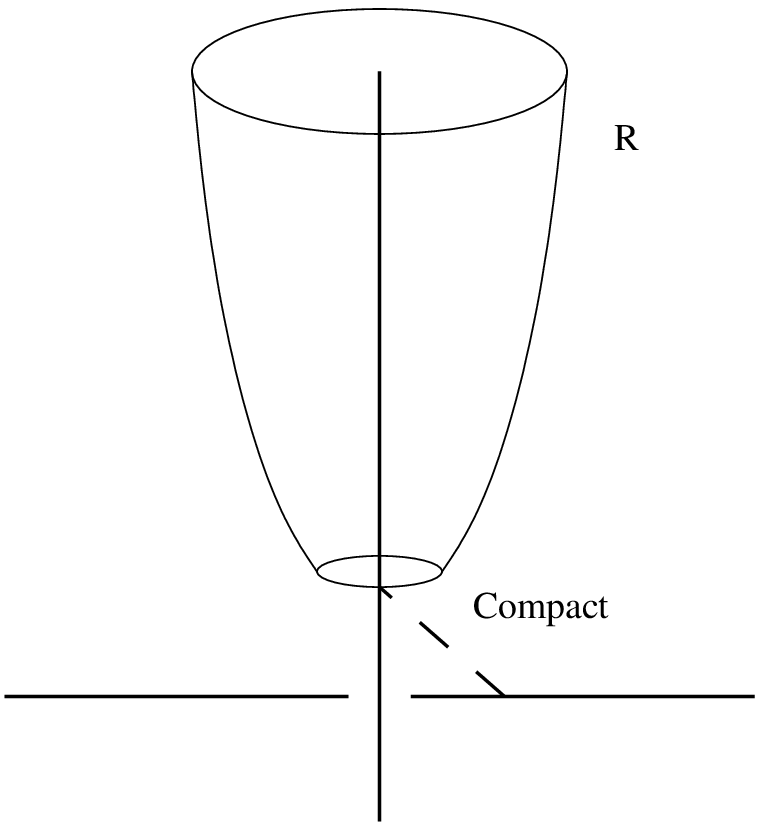}
&
\hspace{0mm}
&
\psfrag{Non-compact}{$P$}
\psfrag{Compact}{}
\includegraphics[scale=.45]{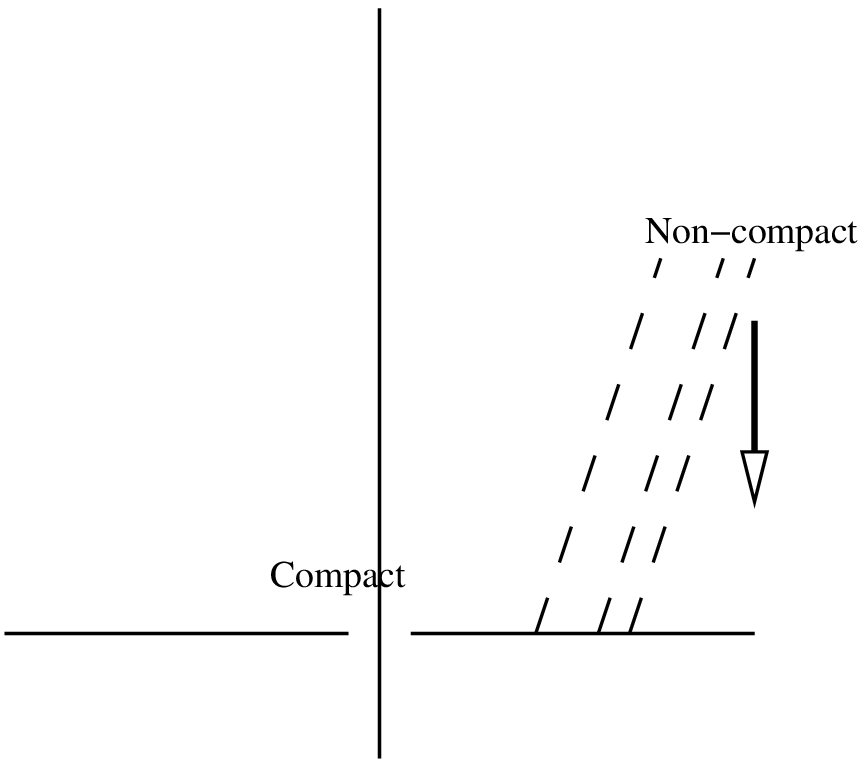}
&
\hspace{0mm}
\\
(a)&&(b)
\end{tabular}
\caption{
(a) The configuration ${\rm F1}_R$ of
multi-wrapped strings inserts the Wilson loop in $R$ into the $U(P)$
Chern-Simons theory on $P$ D-branes.
(b) The compact D-branes develop a ``spike'' and become non-compact.
Strings dissolve into holonomy.}
\label{bion-figure}
\end{figure}

Let us now turn to topological string theory
on the deformed conifold with
 $P$ D-branes  wrapping the $S^3$.
We consider fundamental strings  ending on the branes
along the unknot, and assume that they are
in the multi-string state ${\rm F1}_R$
defined in the previous section.
Recall that $R$ has $P$ rows.
The system is shown in figure \ref{bion-figure}a, where the 
two solid lines are the degeneration loci of the $T^2$ fibers \cite{Aganagic:2002qg}.
We denote by $\alpha$ and $\beta$ the generators of 1-cycles,
and they degenerate along the horizontal and vetical lines, respectively.
We propose that the strings together with the compact D-branes 
are dual to $P$ non-compact
branes.
The $P$ D-branes carry holonomies
\ba
\hat y_i\equiv 
g_s\left (R_i-i+\half(P+N+1)\right),~i=1,\ldots, P.
\ea
In the transition the D-branes develop ``spikes'' and become non-compact,
while the fundamental strings dissolve into the holonomies,\footnote{
It is interesting to note the pattern: the role of fluxes in physical string theory
is played by gauge fields in topological string theory.
In geometric transition, instead of RR flux the complexified \Kahler form
is supported by the grown cycle.
In a topological BIon configuration, the new cycle in the brane world-volume
supports the holonomy rather than field strength.
} as shown in figure \ref{bion-figure}b.
The holonomies tell us where we locate the branes.

The evidence for this duality is the following observation.
The reasoning in the previous section
leads us to define the partition function for the strings
and compact branes
to be the vev of the Wilson loop in the representation $R$.
If the loop is the unknot, the vev is the modular $S$-matrix element $S_{0R}^{(P)}(q)$
for the $U(P)$ current algebra.
The part captured by perturbation theory is
\ba
S^{(P)}_{0R}\sim \prod_{1\leq i<j\leq P} (1-q^{R_i-R_j-i+j})=\prod_{1\leq i<j\leq P} (1-e^{-(\hat y_i-\hat y_j)}).
\ea
The RHS is the contribution of annulus diagrams between 
the non-compact D-branes \cite{Saulina:2004da}.
Since there are no other non-trivial world-sheet instantons, this is precisely
the partition function of $P$ non-compact D-branes with holonomy $\hat U_R$.

\begin{figure}[ht]
\centering
\begin{tabular}{cccccc}
\psfrag{Compact}{$N$}
\psfrag{R}{${\rm F1}_R$}
\includegraphics[scale=.45]{st-br-wil4.eps}
&
\hspace{0mm}
&
\psfrag{Compact}{$N-P$}
\psfrag{Non-compact}{$P$}
\includegraphics[scale=.45]{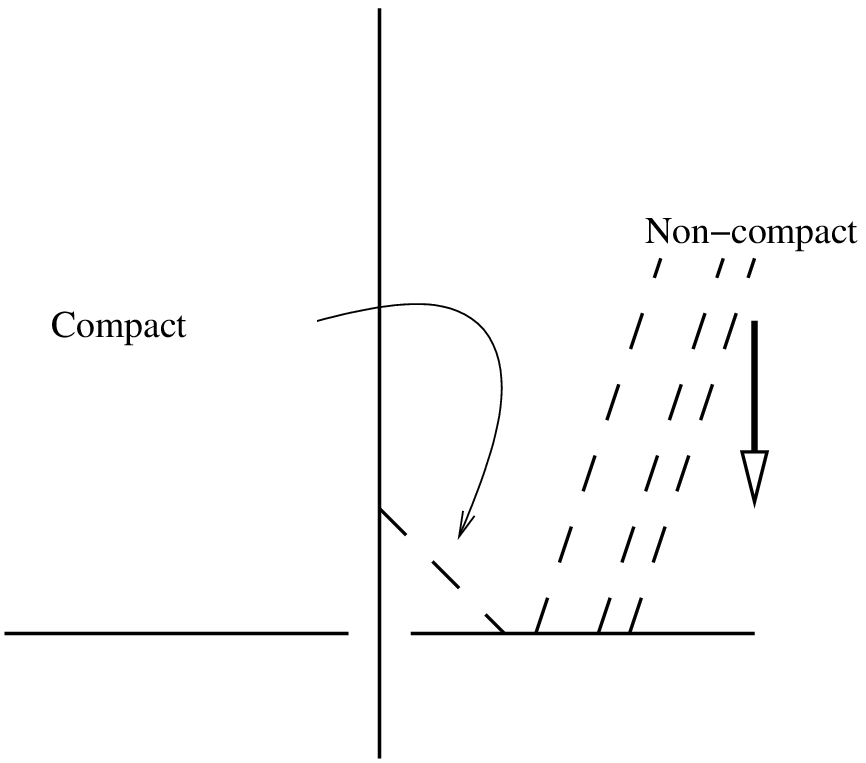}
&
\hspace{0mm}
&
\psfrag{Compact}{$N+M$}
\psfrag{Non-compact}{$\bar M$}
\includegraphics[scale=.45]{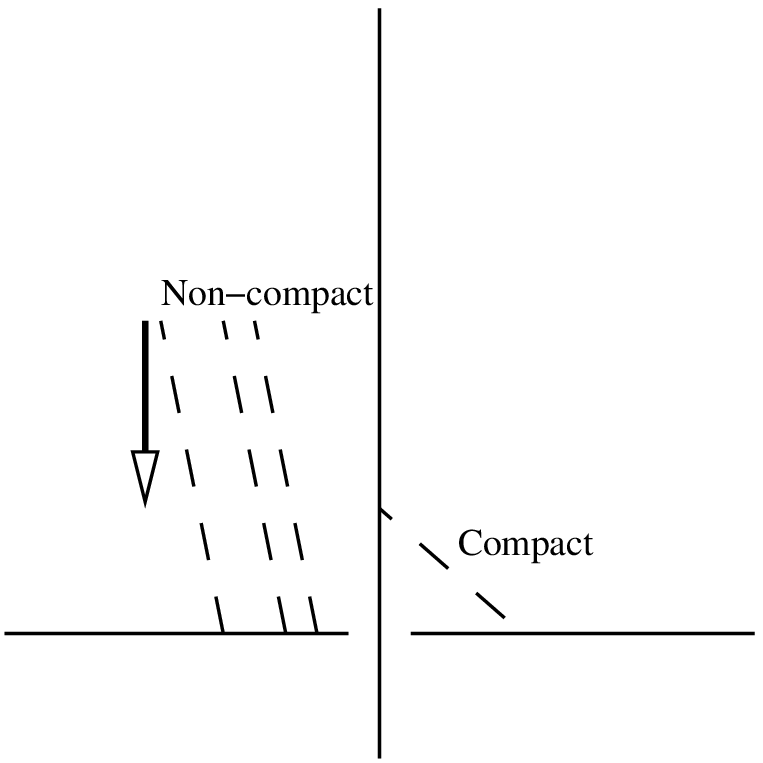}
&
\\
(a)&&(b)&&(c)
\end{tabular}
\caption{
(a) The configuration of
multi-wrapped strings, specified
by $R$, inserts the Wilson loop in $R$ into the $U(N)$
Chern-Simons theory on $N$ D-branes.
(b) $P$ out of $N$ compact D-branes develop a ``spike'' and become non-compact.
(c) The fundamental strings disappear. $M$ new compact D-branes and $M$ non-compact
anti-branes are pair-created.}
\label{string-brane-wilson-deformed}
\end{figure}

As a slight generalization,
we consider $N(>P)$ D-branes on the 
$S^3$ of the deformed conifold and
let fundamental strings ${\rm F1}_R$ end on them as in figure \ref{string-brane-wilson-deformed}a.
We claim  that this system is dual to $P$ non-compact D-branes 
plus $N-P$ compact D-branes
(figure \ref{string-brane-wilson-deformed}b).
The partition function for the strings plus the $N$ D-branes
is the $S$-matrix element $S_{0R}^{(N)}(q)$
for the $U(N)$ current algebra.
We can show the equality\footnote{
See (\ref{loop-brane-amp2}) and (\ref{S-RC}) for derivation.
}
\ba
S_{0R}^{(N)}(q)
\sim
\prod_{1\leq i<j\leq P} (1-e^{-(\hat y_i-\hat y_j)}) \sum_Q
S_{0Q}^{(N-P)}(q)
\Tr_Q\hspace{1mm}{\rm diag}(e^{-\hat y_i})_{i=1}^P,
\label{bion-evidence}
\ea
which we interpret as follows\footnote{
\label{convention-reason}
Making sense of this interpretation
motivated the new convention for D-brane/anti-brane.  
}.
The LHS is the partition function of $N$ compact D-branes with
a configuration of fundamental strings that insert the
Wilson loop in the representation $R$.
These fundamental strings turn into non-compact D-branes
and at the same time strip off $P$ out of $N$
 compact branes.
The factor
$S_{0Q}^{(N-P)}(q)$ is the contribution from the remaining $N-P$ 
compact D-branes.
The $P$ non-compact D-branes
with holonomies $q^{R_i-i+\half(P+N+1)}$ ($i=1,\ldots,P$)
contribute the rest.
The sum is over annulus diagrams between compact and 
non-compact D-branes.\footnote{
The decrease in the number of compact D-branes explains the shift
in the \Kahler modulus of the large $N$ dual resolved conifold found in \cite{Gomis:2006mv}.
}

Similarly, we can show
the identity
\ba
S_{0R}^{(N)}(q)
\sim \hspace{-3mm}
\prod_{1\leq i<j\leq M}
(1-e^{-(x_i-x_j)}) 
\sum_Q
S_{0Q}^{(N+M)}(q) (-1)^{|Q|}
\Tr_{Q^T}{\rm diag}(q^{-R^T_i+i+\half(N-M-1)})_{i=1}^M.\nn\\
\ea
by using the results from \cite{Okuda:2004mb,Gomis:2006mv}.
The RHS is the partition function for $N+M$ compact D-branes
and $M$ non-compact anti-branes with holonomies
$q^{-R^T_i+i+\half(N-M-1)}$ ($i=1,\ldots,M$).
Through the transition, $N$ compact D-branes with strings have turned into
$N+M$ compact D-branes and $M$ non-compact anti-branes, as shown 
in figure \ref{string-brane-wilson-deformed}c.\footnote{
The increase in the number of compact D-branes 
explains the shift
in the \Kahler modulus of the large $N$ dual resolved conifold found 
in \cite{Okuda:2004mb,Gomis:2006mv}.
}

We now generalize the geometry where the BIons sit.
We consider a non-compact Calabi-Yau geometry with the structure
of $T^2\times \R$ fibered over $\R^3$ \cite{Aganagic:2002qg,Marino:2005sj}.
Such a  geometry is more general than toric Calabi-Yau manifolds, and
the deformed conifold is a basic example.
The geometry is specified by a web diagram that is a
 generalization of the toric 
diagram \cite{Aganagic:2002qg}.

\begin{figure}[ht]
\begin{tabular}{ccccc}
\psfrag{R}{\small{${\rm F1}_R$}}
\psfrag{Branes}{$N$}
\psfrag{t2}{$t_2$}
\psfrag{t3}{\small{$t_1$}}
\includegraphics[scale=.43]{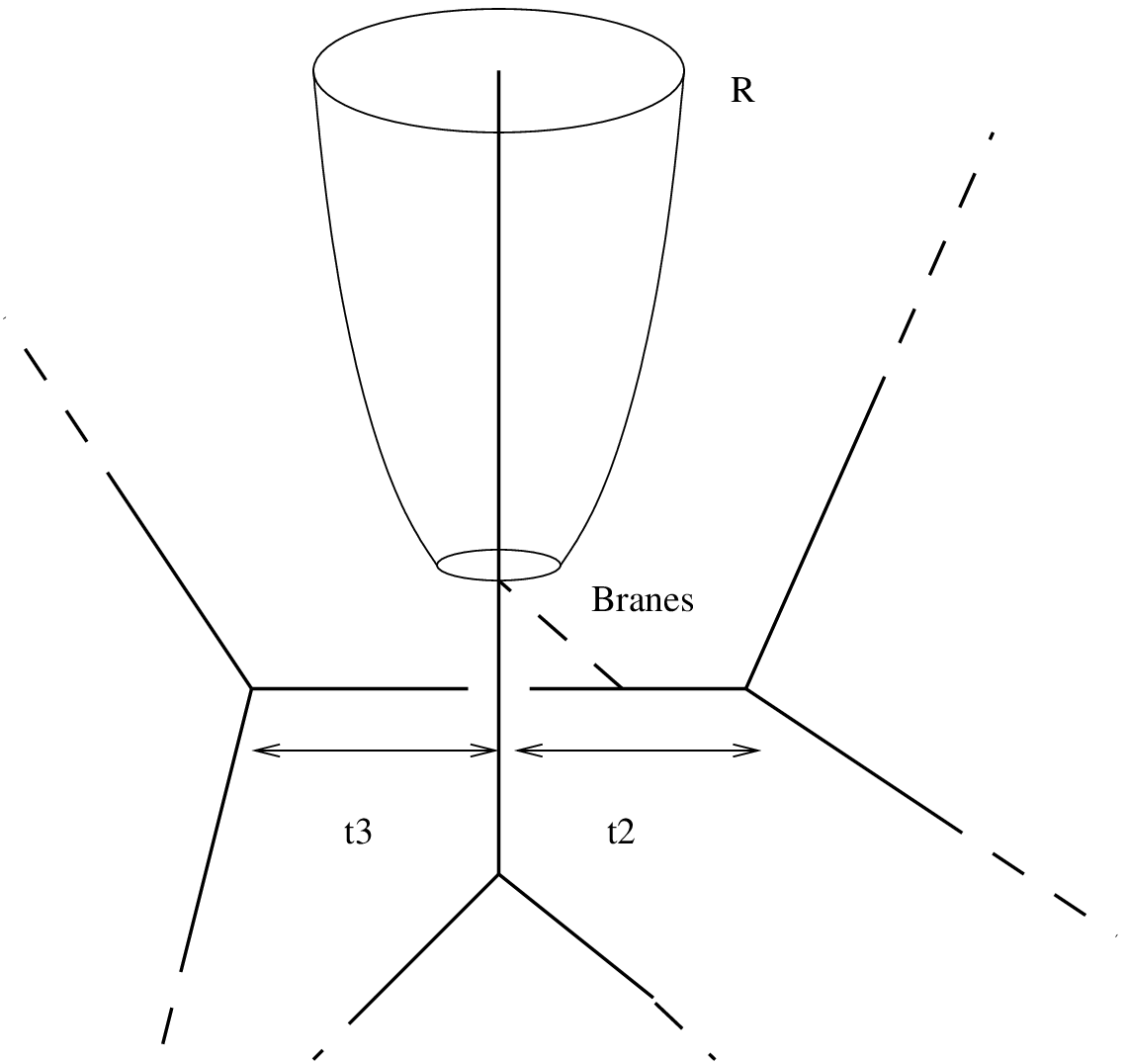}
\hspace{-8mm}
&
\psfrag{Compact}{\small{$N-P$}}
\psfrag{Non-compact}{$P $ }
\psfrag{t2}{\small{$t'_2$}}
\psfrag{t3}{\small{$t'_1$}}
\includegraphics[scale=.43]{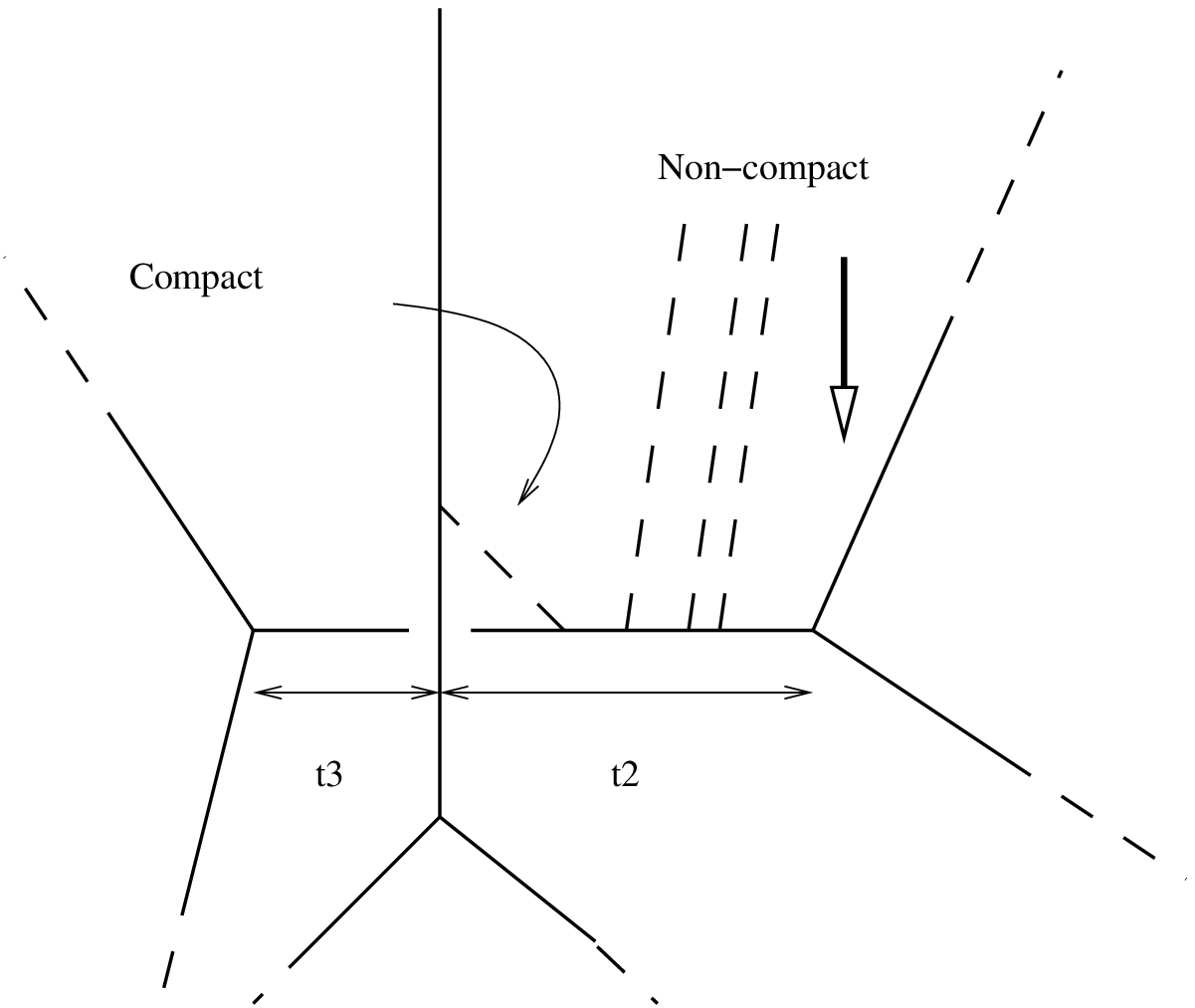}
\hspace{-8mm}
&
\psfrag{Non-compact}{$\bar{M} $ }
\psfrag{Compact}{$N+M$ }
\psfrag{t2}{\small{$\tilde t_2$}}
\psfrag{t3}{\small{$\tilde t_1$}}
\includegraphics[scale=.43]{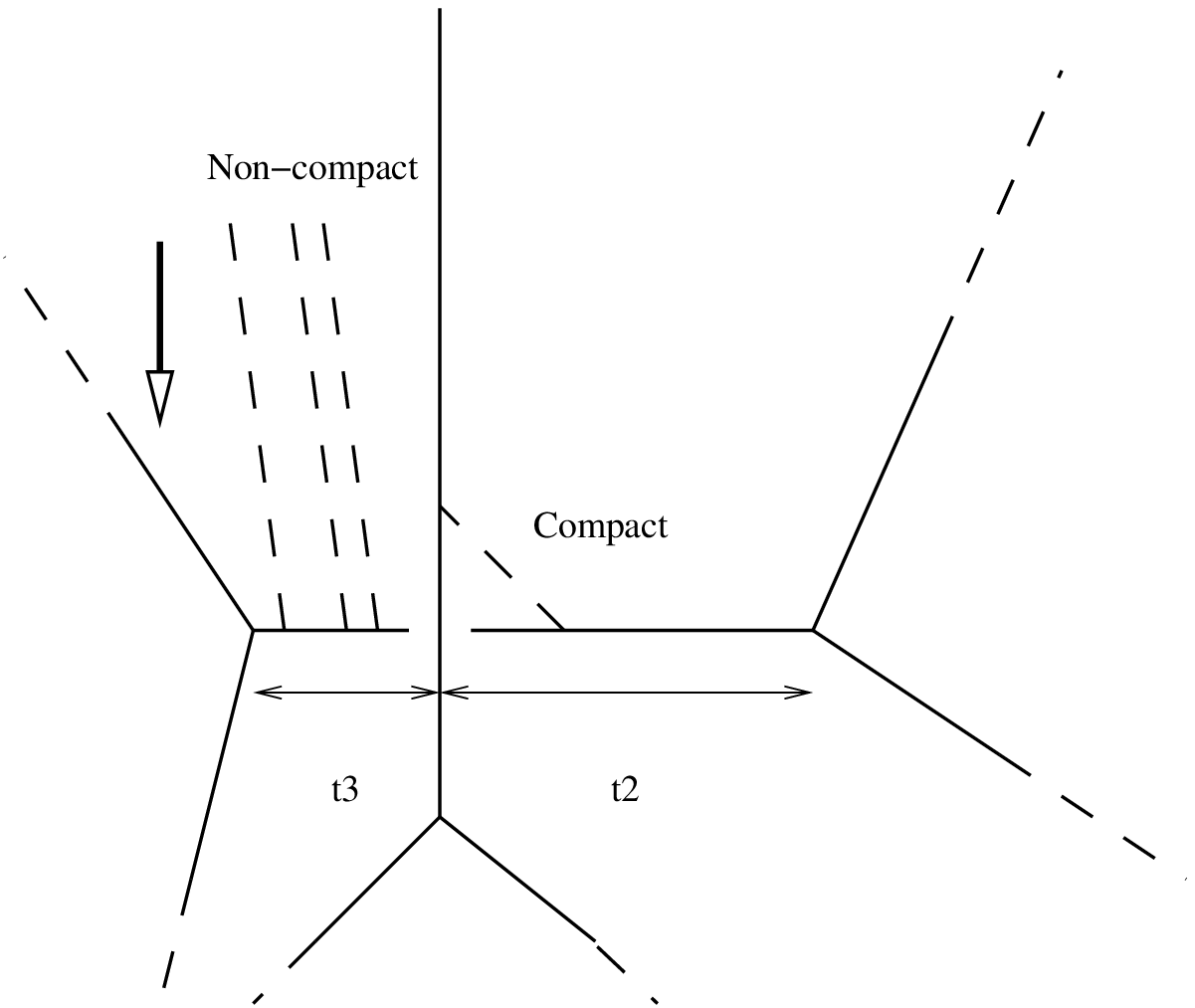}
\\
(a)&(b)&(c)
\end{tabular}
\caption{
(a) A stack of string world-sheets ending on $N$ compact D-branes wrapping $S^3$.
(b)
The strings ending on compact D-branes undergo BIon transition.
$P$ out of $N$ compact D-branes develop ``spikes'' and become non-compact.
The \Kahler moduli are given by $t'_1=t_1-\half g_s P,~t'_2=t_2+\half g_sP$.
(c) In another dual description the strings disappear and $M$ compact D-branes and $M$ non-compact anti-branes
are pair-created.
The \Kahler moduli are given by $\tilde t_1=t_1-\half g_s M, \tilde t_2=t_2+\half g_s M$.
}
\label{string-brane-deformed}
\end{figure}

We assume that  two lines of different orientations cross,
and that one of them is semi-infinite.
By fibering $T^2$ along a segment connecting the two lines, we get a $S^3$.
Let us wrap $N$ D-branes around this $S^3$ as in figure \ref{string-brane-deformed}a.
The local geometry near the $S^3$ the deformed conifold.
Let us consider
a configuration ${\rm F1}_R$ of string world-sheets 
along the semi-infinite line.
The large $N$ dual of the system has the resolved conifold
as local geometry, and 
is a special case
of the situations considered in subsection \ref{st-br-toric-section} \cite{Aganagic:2002qg}.
We simply reinterpret the results there in terms of compact branes.

First, the strings ending on compact D-branes are dual to
the system where
$P$ out of $N$ compact D-branes become non-compact
as shown in figure \ref{string-brane-deformed}b.
 The non-compact D-branes have a particular framing indicated by the arrow
in the figure.
Second, the strings on compact branes are dual to the system where
the strings have disappeared 
and $M$ new compact D-branes and $M$ non-compact anti-branes
are pair-created.
We thus have $N+M$ compact D-branes in total.
This is shown in figure \ref{string-brane-deformed}c
together with the framing of the anti-branes.
Precise values of \Kahler moduli receive well-known non-trivial shifts \cite{Diaconescu:2002sf},
and they can be worked out by combining the results in section \ref{st-br-toric-section}
and the results in the conifold case.
The values are also indicated in the 
figures.

Again we have focused on the quantitative
evidence.
Our BIon proposal
is the large $N$ dual reinterpretation of the string/brane transitions
discussed in the previous section.
We now turn to the physical explanation of the transitions.

\section{String/brane transition and the Fourier transform
} \label{alternative-section}
 


In the BIon solution of physical string theory,
the spike on the world-volume 
is in the direction of the original strings.
However, in the topological string examples of string/brane
transition there
is no obvious sense that the brane world-volume asymptotes 
to the string world-sheet.
In this regard it is natural to introduce additional branes on which
the strings end on.

\begin{figure}[ht]
\centering
\begin{tabular}{cccccc}
\psfrag{a}{$\alpha$}
\psfrag{b}{$\beta$}
\psfrag{-a-b}{$-\alpha-\beta$}
\includegraphics[scale=.45]{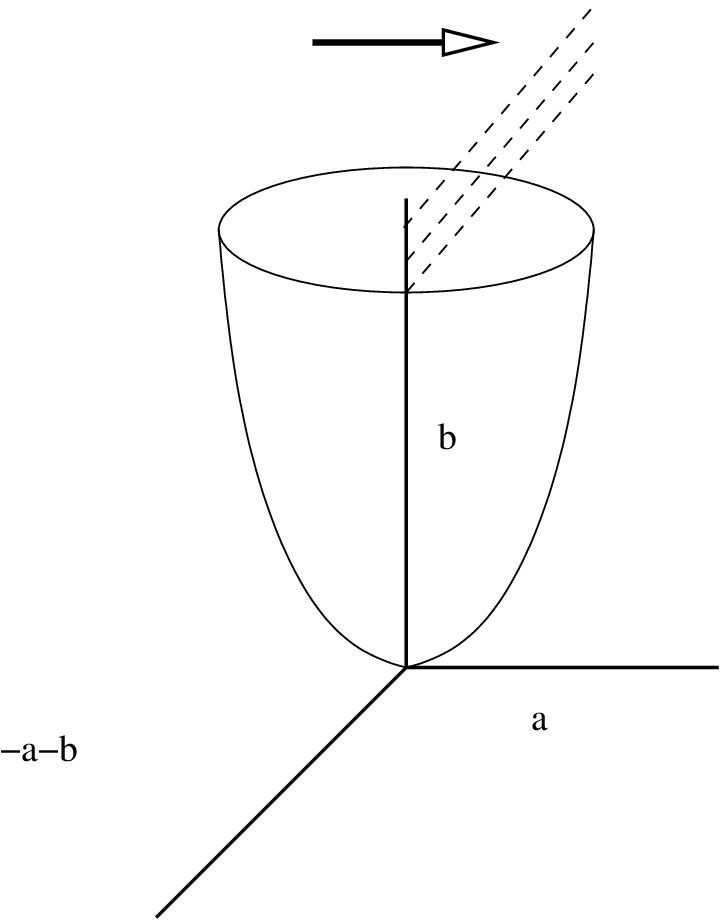}
&
\hspace{10mm}
&
\includegraphics[scale=.45]{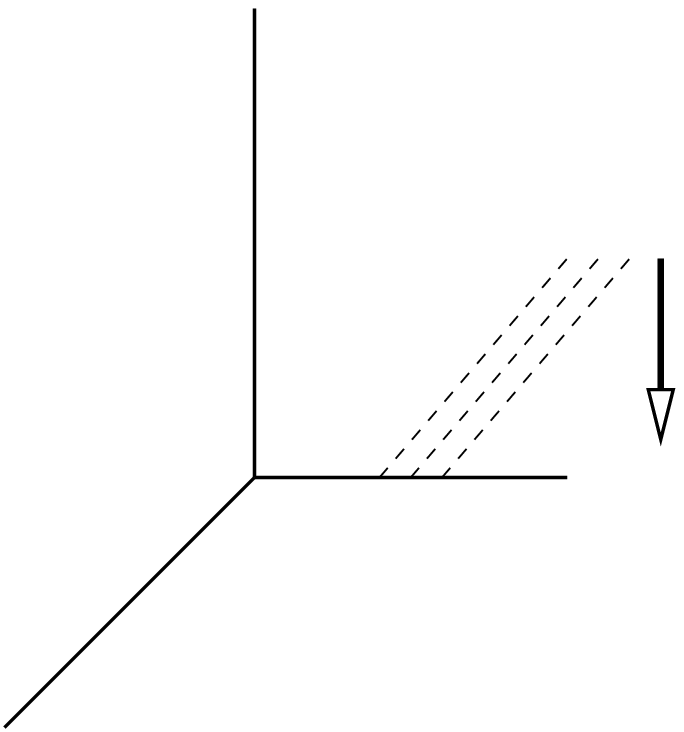}
\\
(a)&&(b)
\end{tabular}
\caption{
(a) For ${\rm C}^3$, 
the 1-cycles $\alpha,\beta$, and $-\alpha-\beta$ of the $T^2$ fiber
degenerate along the right, upper, and lower-left edges.
There are $P$ D-branes on the upper edge.
We impose the asymptotic boundary condition the gauge fields
on the branes,
and it picks out the configuration ${\rm F1}_R$ of  fundamental strings
ending on the branes along the 1-cycle $\alpha$.
(b) The boundary condition is also consistent with
the branes
 on the right edge
with holonomy $\hat U_R$.
}
\label{C3}
\end{figure}

For simplicity we focus on ${\rm C}^3$,
which is the local geometry of any toric Calabi-Yau.
The geometry ${\rm C}^3$
has a $T^2$-fibration structure
\cite{Aganagic:2003db}.
When we introduce non-compact D-branes, the $T^2$ is identified
with the asymptotic boundary of the world-volume.
Let us denote by $\alpha$ and $\beta$ two 1-cycles that generate $H_2(T^2)$.
The cycles $\alpha$, $\beta$, and $-\alpha-\beta$ degenerate along the  edges
in the web diagram as shown in figure \ref{C3}a.

We place $P$ non-compact D-branes on the upper edge.
The world-volume has topology ${\rm R}^2\times S^1$, which we regard as
solid torus.
Since the world-volume has a boundary, we need to impose
a boundary condition on the gauge fields on the branes.
The boundary condition is a state in the Hilbert space
of the Chern-Simons theory on $T^2$,
and the D-brane amplitude is a wave function.
Let us canonically quantize the theory by taking
$\oint_\alpha A$ as coordinates and $\oint_\beta A$ as conjugate momenta.
We denote by $(\alpha,\beta)$ the polarization (choice of conjugate variables),
which is equivalent to the framing  specified by
the horizontal arrow in figure \ref{C3}a  \cite{Aganagic:2003db}.
As a basis of the Hilbert space let us take
the coordinate eigenstates ${}_{(\alpha,\beta)}\langle V|$
such that
\ba
{}_{(\alpha,\beta)}\langle V|{\rm P}e^{-\oint_{\alpha} A}
={}_{(\alpha,\beta)}\langle V|V,
\ea
where $V$ is a $P\times P$ matrix.
The D-brane amplitude $Z_{(\alpha,\beta)}(V)$ 
is the wave function
\ba
Z_{(\alpha,\beta)}(V)={}_{(\alpha,\beta)}\langle V|Z\rangle,
\ea
where $|Z\rangle$ is the state in the Hilbert space
intrinsically defined by topological string theory in the given background.
The amplitude $Z^{(\alpha,\beta)}(V)$ 
is computed by summing over many possible string
configurations that end on the D-branes
while fixing the background holonomy $V$ along $\alpha$.
This is the conventional treatment of non-compact D-branes \cite{Ooguri:1999bv,Aganagic:2003db}.

Another important basis consists of states ${}_{(\alpha,\beta)}\langle R|$.
When the state
${}_{(\alpha,\beta)}\langle R|$ is 
used as a boundary condition on non-compact branes,
it picks out the configuration ${\rm F1}_R$ of strings.
This is the defining property of the state.
The Fourier transform
\ba
Z_{(\alpha,\beta)}(V)=\sum_R \Tr_R V \ Z_{(\alpha,\beta),R}
\ea
relates the D-brane amplitude $Z_{(\alpha,\beta)}(V)$ to 
\ba
Z_{(\alpha,\beta),R}\equiv{}_{(\alpha,\beta)}\langle R|Z\rangle,
\ea
which is the partition function of strings in the configuration ${\rm F1}_R$
as we defined in section \ref{st-br-toric-section}.

Suppose we use the state ${}_{(\alpha,\beta)}\langle R|$ 
as a boundary condition for the $P$ D-branes.
It was shown in \cite{Elitzur:1989nr} that 
${}_{(\alpha,\beta)}\langle R|$ is in fact a 
 momentum eigenstate:
\ba
{}_{(\alpha,\beta)}\langle R|
\Path e^{-\oint_\beta A}
=
{}_{(\alpha,\beta)}\langle R|
\hat U_R,~~~\hat U_R\equiv{\rm diag}(q^{R_i-i+1/2+P/2}).
\ea
On the upper edge,
$\beta$ is a contractible cycle and cannot support a non-trivial holonomy.
To interpret the non-trivial holonomy, we note that
the cycle $\beta$ is non-contractible on the right edge.
Indeed, $(\beta,-\alpha)$ is precisely the polarization
for the D-branes on the right edge with framing specified by the vertical arrow as shown in
figure \ref{C3}b.
Therefore we have that
\ba
{}_{(\alpha,\beta)}\langle R| \propto {}_{(\beta,-\alpha)}\langle \hat U_R|,
\ea
and the partition function
$Z_{(\alpha,\beta),R}$ for strings
should be identified with the partition function
of the D-branes
on the right edge with holonomy $\hat U_R$ and the specified framing:
\ba
Z_{(\alpha,\beta),R}\propto Z_{(\beta,-\alpha)}(\hat U_R).
\ea
This is exactly what we found in section \ref{st-br-toric-section},
and explains the string/brane transition.

\begin{figure}[ht]
\centering
\begin{tabular}{cccccc}
\psfrag{a}{$\alpha$}
\psfrag{b}{$\beta$}
\psfrag{-a-b}{$-\alpha-\beta$}
\includegraphics[scale=.45]{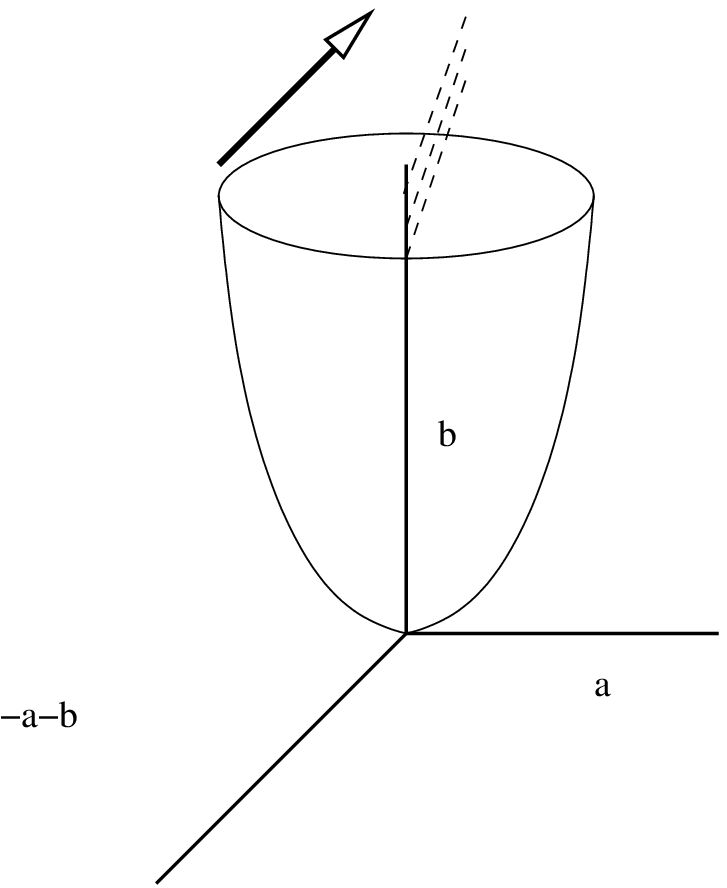}
&
\hspace{10mm}
&
\includegraphics[scale=.45]{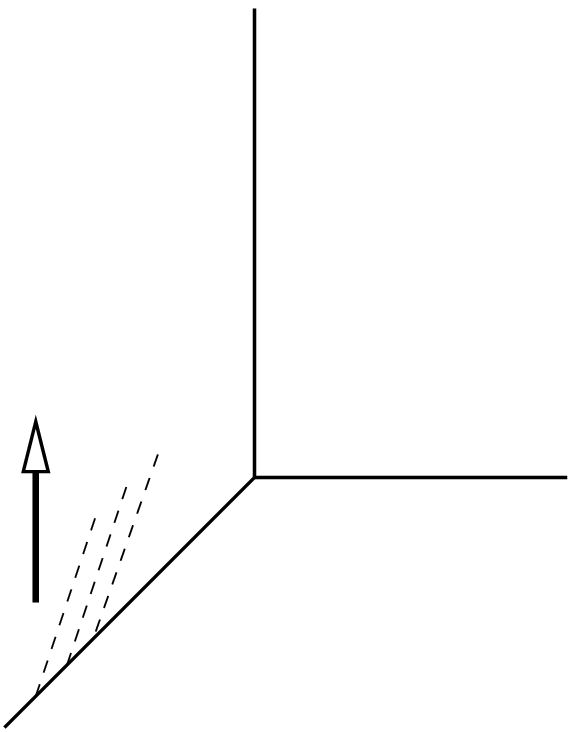}
\\
(a)&&(b)
\end{tabular}
\caption{
(a)
The strings extended along the upper edge end on anti-branes along $\alpha$.
(b) The boundary condition imposed on the anti-branes
that pick out the configuration ${\rm F1}_R$ of  fundamental strings
is equivalent to the condition that the anti-branes end on the lower-left edge
with holonomy $\hat U_{R^T}\equiv {\rm diag}(q^{R^T_i-i+1/2+M/2})_{i=1}^M$.
}
\label{C3-anti}
\end{figure}
We can also understand the transition of strings to anti-branes.
Let us begin with the anti-branes with the framing specified by the arrow  shown in figure \ref{C3-anti}a.
This framing corresponds to taking $\oint_{\alpha+\beta} A$ and $\oint_\beta A$
as canonical coordinates and momenta, respectively.
The partition function can be expanded as
\ba
\bar Z_{(\alpha+\beta,\beta)}(V)=\sum_R \Tr_R V \ {}_{(\alpha+\beta,\beta)}\langle R|\bar Z\rangle.
\ea
Since anti-branes are related to D-branes by substitution
$\Tr_R V\ra (-1)^{|R|}\Tr_{R^T} V$ \cite{Aganagic:2003db},
we have that $|\bar Z\rangle=C|Z\rangle$, where $C$ is the charge conjugation
operator defined in Appendix \ref{rewriteoperator}.  Thus   we find that
\ba
Z_{(\alpha+\beta,\beta),R}&=&{}_{(\alpha+\beta,\beta)}\langle R|Z\rangle\nn\\
&=&{}_{(\alpha+\beta,\beta)}\langle R|CC|Z\rangle\nn\\
&=&(-1)^{|R|}{}_{(\alpha+\beta,\beta)}\langle R^T|\bar{Z}\rangle\nn\\
&\propto&{}_{(\beta,-\alpha-\beta)}\langle \hat U_{R^T}|\bar Z\rangle\nn\\
&=&\bar Z_{(\beta,-\alpha-\beta)}(\hat U_{R^T}).
\ea
The matrix $\hat U_{R^T}$ is the holonomy along $\beta$, which is 
contractible on the upper edge
but  non-contractible on the lower-left edge.
Moreover, $(\beta,-\alpha-\beta)$ is the polarization for Chern-Simons theory
on the anti-branes sitting at the lower-left edge with the framing given by
the arrow in figure \ref{C3-anti}b.
This argument implies that $Z_{(\alpha+\beta,\beta),R}$ is proportional
to the partition function of the anti-branes with holonomy $\hat U_{R^T}$,
just as we found in section \ref{st-br-toric-section}.

\section*{Acknowledgments}
I thank Jaume Gomis for collaboration on related projects and for reading the manuscript.
I'm also grateful to Dan Jafferis and Kentaroh Yoshida for useful correspondence and discussion.
My research is supported in part by the NSF grants PHY-05-51164 and PHY-04-56556.
 \appendix

\renewcommand{\theequation}{\Alph{section}\mbox{.}\arabic{equation}}

 \bigskip\bigskip
 \noindent {\LARGE \bf Appendix}


\section{Unknot Wilson loop vev as a brane amplitude}\label{rederivation}

In this appendix we will rederive the relation between
the unknot Wilson loop vev and 
the partition function of D-branes in the resolved conifold.
The original derivation in \cite{Gomis:2006mv} used
a Calabi-Yau crystal in an intermediate
step.
Here we present a more elementary 
derivation.
We will also formulate a conjecture
relating knot polynomials and D-brane amplitudes
for general knots.

%

For the canonically framed unknot, the normalized  vev $W_R^{(N)}(q)$
in $U(N)$ Chern-Simons theory is 
\ba
&&W_R^{(N)}(q)=S^{(N)}_{0R}/S^{(N)}_{00}=\Tr_R\hspace{1mm} {\rm diag}(q^{-i+1/2+N/2})_{i=1}^N.
\ea
We manipulate this expression as
\ba
\nn\\
&&\f{\det_{1\leq i,j\leq N}(q^{(-i+1/2+N/2)(R_j-j+N)})}{
\det_{1\leq i,j\leq N}(q^{(-i+1/2+N/2)(-j+N)})}
\nn\\
& =&
q^{-\half(N-1)|R|}\f{\det_{1\leq i,j\leq N}(q^{(-i+N)(R_j-j+N)})}{
\det_{1\leq i,j\leq N}(q^{(-i+N)(-j+N)})}\nn\\
&=&
q^{-\half(N-1)|R|}\f{\prod_{i<j\leq N}\left(q^{R_i-i+N}-q^{R_j-j+N}\right)}
{\prod_{i<j\leq N}\left(q^{-i+N}-q^{-j+N}\right)}
\nn
\ea
\ba
&=&
q^{\sum_j (j-\half -\half N)R_j}\prod_{i<j\leq N}\f{1-q^{j-i+R_i-R_j}}
{1-q^{j-i}}
\nn
\\
&=&
q^{\sum_j (j-\half -\half N)R_j}\prod_{i<j\leq P}\f{1-q^{j-i+R_i-R_j}}
{1-q^{j-i}}\prod_{i\leq P<j\leq N} \f{1-q^{j-i+R_i}}{1-q^{j-i}}
\nn
\\&=&
q^{\sum_j (j-\half -\half N)R_j}\left[\prod_{i<j\leq P}(1-q^{j-i+R_i-R_j})\right]
\left[\prod_{i<j\leq P}\oo{1-q^{j-i}}\right]\nn\\
&&~~~\times
\left[\prod_{i=1}^P\prod_{j=1}^{N-P}(1-q^{P+j-i+R_i})\right]
\left[\prod_{i=1}^P\prod_{j=1}^{N-P}
\oo{1-q^{P+j-i}}\right].\label{manipulation}
\ea
The annulus diagrams between D-branes $\prod_{i<j\leq P} (1-e^{-(y_i-y_j)})$
contribute the first angular bracket in the last expression.
The third represents the D-brane amplitudes:
\ba
\prod_{i=1}^P\prod_{j=1}^{N-P}(1-q^{P+j-i+R_i})
&=&\prod_{i=1}^P\prod_{j=1}^\infty \f{1-q^{P+j-i+R_i}}{1-q^{N+j-i+R_i}}\nn\\
&=&\prod_{i=1}^P\exp \sum_{n=1}^\infty \f{e^{-n y_i}-e^{-n(\tilde t+y_i)}}{n[n]}.
\ea
Here we have defined the shifted \Kahler modulus $\tilde t=g_s(N-P)$.
The product of the second and the fourth brackets in (\ref{manipulation})
is
the ratio of closed string amplitudes with different values of moduli:
\ba
&&
\left[\prod_{i<j\leq P}\oo{1-q^{j-i}}\right]
\left[\prod_{i=1}^P\prod_{j=1}^{N-P}
\oo{1-q^{P+j-i}}\right]=\left[\prod_{j=1}^P \oo{(1-q^j)^{P-j}}
\right]\left[\prod_{i=1}^P\prod_{j=1}^{N-P}
\oo{1-q^{i+j-1}}\right]\nn\\
&=&
\left[
\prod_{j=1}^P\f{(1-q^j)^j}{(1-q^j)^P}
\right]
\left[\prod_{i=1}^\infty\prod_{j=1}^\infty
\f{1-q^{P+i+j-1}}{1-q^{i+j-1}}\f{1-q^{N-P+i+j-1}}{1-q^{N+i+j-1}}\right]\nn\\
&=&
\left[
\prod_{j=1}^\infty\f{(1-q^j)^j(1-q^{P+j})^P}{(1-q^j)^P (1-q^{P+j})^{P+j}}
\right]
\left[
\prod_{j=1}^\infty\f{(1-q^{P+j})^j(1-q^{N-P+j})^j}{(1-q^j)^j(1-q^{N+j})^j}\right]\nn\\
&=&
\xi(q)^P
\prod_{j=1}^\infty
\f{(1-q^{N-P+j})^j}{(1-q^{N+j})^j}
=\xi(q)^P
\exp\left(\sum_{n=1}^\infty \f{e^{-n  t}}{n[n]^2}-\sum_{n=1}^\infty \f{e^{-n 
\tilde t}}{n[n]^2}\right).
\ea
To summarize, we have found that
\ba
\left[\exp- \sum_{n=1}^\infty \f{e^{-n t}}{n[n]^2}\right] W^{(N)}_R(q)
=\left[ q^{\sum_j (j-\half -\half N)R_j}\xi(q)^P\prod_{i<j\leq P}(1-e^{-(y_i-y_j)})\right]\nn\\
\times
\left[\prod_{i=1}^P\exp \sum_{n=1}^\infty \f{e^{-n y_i}-e^{-n(\tilde t+y_i)}}{n[n]}\right]
\left[\exp-\sum_{n=1}^\infty \f{e^{-n \tilde t}}{n[n]^2}\right]. \label{loop-brane-amp1}
\ea

Let us express (\ref{loop-brane-amp1}) in a new form,
where
the precise expression of the D-brane amplitude does not appear.
We define the generating functionals, or Fourier transforms, 
of Wilson loop vevs \cite{Ooguri:1999bv}:
\ba
G_+^{(N)}(q,V)&:=&\sum_R W^{(N)}_R (q) \Tr_{R} V,\\
G_-^{(N)}(q,V)&:=&\sum_R W^{(N)}_R (q) (-1)^{|R|}\Tr_{R^T} V.
\ea
For the unknot we have
\ba
G_\pm^{(N)}(q,V)=\prod_{i=1}^N \det(1 - q^{i-(N+1)/2}V)^{\mp 1}=
\exp \left(\pm \sum_{n=1}^\infty\f{e^{\half g_sN}-e^{-\half g_s N}}{n[n]}\right)\Tr V^n.
\ea
Thus (\ref{loop-brane-amp1}) can be written as
\ba
 W^{(N)}_R(q)
= N_R\
 G_+^{(N-P)}(q, e^{-\tilde t/2}U_R), \label{loop-brane-amp2}
\ea
where we have defined the knot-independent factor
\ba
N_R\equiv q^{\sum_j (j-\half -\half N)R_j}\xi(q)^P 
\left[\prod_{i<j\leq P}(1-e^{-(y_i-y_j)})\right]
\exp \left(\sum_{n=1}^\infty \f{e^{-n \tilde t}-e^{-n t}}{n[n]^2}\right).
\ea
Up to $N_R$,
(\ref{loop-brane-amp2}) tells us 
that {\it the unknot Wilson loop vev
is the Fourier transform of its own.}
The RHS depends on $R$ almost exclusively through $e^{-y_i}$,
just like it depends on $N$ only through $e^{-t}$ and $e^{-\tilde t}$.
Though we derived (\ref{loop-brane-amp2}) for the unknot,
we conjecture that the relation 
between the knot polynomials $W_R(q,\lambda)\equiv W^{(N)}_R(q)$ 
(here $\lambda=q^N$)
and their generating functional($=$ D-brane partition function) 
$G_+(q,\lambda, V)\equiv G_+^{(N)}(q,V)$
should hold more generally 
for any knot with a universal normalization factor $N_R$.
It would be interesting to check the conjecture
for torus knots,
 together with the corresponding relation
involving $G_-$.
The results for torus knots in \cite{Labastida:2000zp} may be useful.

%
To rewrite (\ref{loop-brane-amp2}) in the form of (\ref{bion-evidence}),
note that the Chern-Simons partition function
can be identified with the resolved conifold
partition function  \cite{Gopakumar:1998ki}:
\ba
S_{00}^{(N)}=\left(\f{g_s}{2\pi}\right)^{N/2}
e^{-\f{\pi i}4 N^2}
q^{-N(N-1)/12}
\xi(q)^{-N}
M(q) \exp\left( -\sum_n \f{e^{-nt}}{n[n]^2}\right).\label{S-RC}
\ea
For precise equality, we should multiply the RHS of
 (\ref{bion-evidence}) by the prefactor
that contains genus-zero and -one contributions:
\ba
 \left(\f{g_s}{2\pi}\right)^{P/2}
e^{-({N^2-(N-P)^2})\oo 4 \pi i}
q^{\sum_j (j-\half -\f{N}2)R_j -\f{N(N-1)}{12}+\f{(N-P)(N-P-1)}{12}
} .
\ea
As is often the case for low-genus contributions,
the topological string interpretation of this prefactor
is not clear.

\section{Operator formalism}
\label{rewriteoperator}

Let us review
the relation between the representation theory of $U(N)$ and   two dimensional bosons and fermions in two dimensions.
The formalism is useful
in deriving the group theory identities in the main text
and dealing with canonical quantization of Chern-Simons theory on $T^2$.

Let us consider the mode expansion of a chiral boson $\phi(z)$ and fermions $\psi(z), \bar\psi(z)$ in two dimensions, which are
related by bosonization:
\ba
&&\phi(z)=
i\sum_{n\neq 0} \f{\alpha_n}{nz^{n}},~~
\psi(z)=\sum_{r\in\Z+\half}\f{\psi_r}{z^{r+1/2}},~~
\bar\psi(z)=\sum_{r\in\Z+\half}\f{\bar\psi_r}{z^{r+1/2}},
\\
&&i\p\phi=:\psi\bar\psi:,~~\psi=:e^{i\phi}:,~~\bar\psi=:e^{-i\phi}:.
\label{modeexp}
\ea
The  oscillator modes satisfy the  commutation relations:
\ba
[\alpha_n,\alpha_m]=n\delta_{n+m,0}\qquad \{\psi_r,{\bar\psi_s}\}=\delta_{r+s,0}.
\ea
We can also define a charge conjugation operator $C$.
It 
exchanges $\psi$ and $\bar\psi$:
\ba
C\psi(z)C=\bar\psi(z),~C^2=1,~C|0\rangle=|0\rangle.
\ea
Then $C$ acts on $i\p\phi(z)=:\psi(z)\bar\psi(z):$ as:
\ba
C\p\phi(z)C=-\p\phi(z).
\ea
The connection between Young tableau $R$ and fermions arises from the identification
\ba
|R\rangle=\prod_{i=1}^d \psi_{-a_i-1/2}\bar\psi_{-b_i-1/2}|0\rangle,
\label{fermionstate}
\ea
where $a_i\equiv  R_i-i$,  $b_i=R_i^T-i$ are the Frobenius coordinates of $R$,
 and $d$ is the number of boxes in the diagonal of the Young tableau $R$.
It follows that:
\ba
C|R\rangle=(-1)^{|R|}|R^T\rangle. \label{charge-conjugation}
\ea

Let us now define \cite{Okounkov:2003sp} the operator
\ba
\Gamma_\pm(z):=\exp \sum_{n=1}^\infty \f{z^{\pm n}}{n} \alpha_{\pm
n},
\ea
which  satisfies the relations
\ba
\Gamma_+(z_+)\Gamma_-(z_-)=\oo{1-z_+/z_-}\Gamma_-(z_-)\Gamma_+(z_+),~~
\Gamma_+(z)|0\rangle=|0\rangle,~\langle 0|\Gamma_-(z)=\langle 0|. \label{Gamma+-}
\ea
The skew Schur polynomials can be conveniently expressed as
\ba
s_{R/Q}(x)=\langle R|\prod_i\Gamma_-(x_i^\mo)|Q\rangle=\langle Q|\prod_i\Gamma_+(x_i)|R\rangle.
\label{skewschur}
\ea
The familiar Schur polynomials $s_R(x)\equiv \Tr_R {\rm diag}(x_i)$ 
arise when $|Q\rangle=|0\rangle$. 
In terms of the Schur polynomials,
the skew Schur polynomials are given by
\ba
s_{R/Q}(x)=\sum_{R'}N^R_{QR'}s_{R'}(x), \label{skew}
\ea
where $N^R_{QR'}$ are the Littlewood-Richardson coefficients.


\section{Topological vertex amplitude}\label{top-vert-app}
We use the convention such that $q$ is replaced by $q^\mo$ relative to \cite{Aganagic:2003db}.
Explicitly the topological vertex 
amplitude is given, with slight abuse of notation, by:
\ba
\hspace{-2mm}
C_{R_1R_2R_3}(q)
\hspace{-1mm}
=
\hspace{-1mm}
q^{-\half(\kappa_{R_2}+\kappa_{R_3})}
s_{{R_2}^T}(q^{i-1/2})
\hspace{-1mm}
\sum_Q
\hspace{-1mm}
 s_{R_1/Q}(q^{-(R_2^T)_i+i-1/2})
s_{{R_3}^T/Q}(q^{-(R_2)_i+i-1/2}).
\ea
Here $s_{R_1/R_2}$ is a skew Schur function.  The index $i$ runs from $1$ to $\infty$.

In the next appendix,
we need the crystal representation of the topological vertex.
This is obtained by a simple manipulation of the results in \cite{Okounkov:2003sp}:
\ba
C_{R_1R_2R_3}=
M(q)^\mo q^{-\half \kappa_{R_1}+\half||R^T_3||^2}
\left\langle
R_2\right|\left[\prod_{i=1}^{\infty} \Gamma_\pm(q^{i-1/2})\right]\left[\prod_{i=1}^{\infty}
\Gamma_\pm(q^{-i+1/2})\right]\left|R_1{}^T\right\rangle. \label{vertex-crystal}
\ea
In this expression, the pattern of $\Gamma_\pm$ is determined by
$R_3$ as described in \cite{Okounkov:2003sp},
and the border between the two products is the diagonal of $R_3$.

The partition function of topological strings on any toric Calabi-Yau manifold,
with or without $D$-branes, can be computed by gluing several topological vertices.
The gluing rules are explained in \cite{Aganagic:2003db}.

\section{An identity for string/brane transitions} \label{st-br-identity-section}

The aim of this appendix is to prove the identity (\ref{st-br-identity}).
\ba
&&\sum_{Q_2,Q_2'} C_{\cdot Q_2 R_3}(-1)^{|Q_2|} q^{\half \kappa_{Q_2}}
\Tr_{Q_2/Q_2'} U_R (-1)^{|Q_2'|} \Tr_{R_2{}^T/Q_2'{}^T} U_R{}^\mo
\nn\\
&=& \sum_{Q_2,Q_2'}\langle R_3\left[\prod_{i=1}^\infty \Gamma_-(q^{-i+1/2})\right]
\left[\prod_{i=1}^\infty \Gamma_+(q^{i-1/2})\right] |Q_2{}^T\rangle (-1)^{|Q_2|}
\langle Q_2|
\nn\\
&&\times \left[\prod_{i=1}^P \Gamma_- (q^{-R_i+i-P-1/2})\right] |Q_2'\rangle
(-1)^{|Q_2'|} \langle Q_2'{}^T|\left[
\prod_{i=1}^P \Gamma_+(q^{-R_i+i-P-1/2})\right] |R_2{}^T\rangle
\nn\\
&=& \langle R_3|\left[\prod_{i=1}^\infty \Gamma_-(q^{-i+1/2})\right]
\left[\prod_{i=1}^\infty \Gamma_+(q^{i-1/2})\right]
\nn\\
&&\times C\left[\prod_{i=1}^P \Gamma_- (q^{-R_i+i-P-1/2})\right] C\left[
\prod_{i=1}^P\Gamma_+(q^{-R_i+i-P-1/2})\right] |R_2{}^T\rangle
\nn
\ea
\ba
&=& \langle R_3|\left[\prod_{i=1}^\infty \Gamma_-(q^{-i+1/2})\right]
\left[\prod_{i=1}^\infty \Gamma_+(q^{i-1/2})\right]
\nn\\
&&\times \left[\prod_{i=1}^P \Gamma_-^\mo (q^{-R_i+i-P-1/2})\right]
\left[
\prod_{i=1}^P\Gamma_+(q^{-R_i+i-P-1/2})\right] |R_2{}^T\rangle
\nn\\
&=&
\langle R_3|
\left[ \prod_{i=1}^\infty \Gamma_+(q^{i-1/2})\right]
\left[ \prod_{i=1}^{\ra \infty} \Gamma_-(q^{-i+1/2})\right]
\left[\prod_{i=1}^{ P\la} \Gamma_-^\mo (q^{-R_i+i-P-1/2})\Gamma_+(q^{-R_i+i-P-1/2})\right]
\nn\\
&&\times |R_2{}^T\rangle
M(q)^\mo \prod_{i=1}^P \prod_{j=1}^{i-1}(1-q^{-R_i+i+R_j-j})^\mo
\nn\\
&=&
\langle R_3|
\left[ \prod_{i=1}^\infty \Gamma_\pm(q^{i-1/2})\right]
\left[ \prod_{i=1}^{\ra \infty} \Gamma_\mp(q^{-i+1/2})\right]
|R_2{}^T\rangle\nn\\
&&\times
q^{P|R_3|-P|R_2|}M(q)^\mo \prod_{i=1}^P \prod_{j=1}^{i-1}(1-q^{-R_i+i+R_j-j})^\mo
\left(\prod_{j=1}^\infty(1-q^{j})\right)^P
\nn\\
&=&\xi(q)^{-P} \prod_{1\leq i<j\leq P} (1-e^{-(a_i-a_j)})^\mo C_{RR_2R_3}
q^{\half \kappa_{R_2}-\half ||R^T||^2}q^{P|R_3|-P|R_2|}.
\ea
Here we used the crystal representation (\ref{vertex-crystal}) of the topological vertex.
The arrows above the product symbols indicate the direction we order the factors.

\bibliography{st-br11}

\end{document}